# Temperature Dependence of Electron to Lattice Energy-Transfer in Single-Wall Carbon Nanotube Bundles

G. Moos,[a] R. Fasel,[b] and T. Hertel [a]

[a] Department of Physical Chemistry, Fritz-Haber-Institut der Max-Planck-Gesellschaft, Faradayweg 4-6, D-14195 Berlin, Germany
[b] EMPA Dübendorf, Überlandstr. 129, 8600 Dübendorf, Switzerland

The electron-phonon coupling strength in single-wall carbon nanotube (SWNT) bundles has been studied directly in the time-domain by femtosecond time-resolved photoelectron spectroscopy. We have measured the dependence of $H(T_e,T_l)$, the rate of energy-transfer between the electronic system and the lattice as a function of electron and lattice temperatures $T_e$ and $T_l$. The experiments are consistent with a $T^5$ dependence of $H$ on the electron- and lattice-temperatures, respectively. The results can be related to the *e-ph* mass enhancement parameter $\lambda$. The experimentally obtained value for $1/Q^2_D$, where $Q_D$ is the Debye temperature, suggests that *e-ph* scattering times at the Fermi level of SWNT bundles can be exceptionally long, exceeding 1.5 ps at room temperature.

## 1. INTRODUCTION

If the electronic system of a solid at temperature $T_e$ and its lattice at temperature $T_l$ are brought out of thermal equilibrium, the two subsystems will try to equilibrate by the exchange of energy through electron-phonon (*e-ph*) interactions. The rate at which energy is exchanged naturally depends on the *e-ph* coupling strength and can be related to the electron-phonon mass-enhancement parameter $\lambda$ [1]. The latter is of central importance for the BCS theory of superconductivity and has been used successfully to predict transition temperatures in BCS superconductors [2,3]. However, non-equilibrium electron and lattice temperatures have to be generated artificially, by high field gradients and current densities or by laser heating, for example. Here, we use laser heating via optical absorption for generating non-equilibrium electron- and lattice temperatures. The electronic temperature is probed by monitoring the width of the Fermi distribution at the Fermi level with photoelectron-spectroscopy. In combination with ultrashort laser pulses this technique has become known as femtosecond time-resolved photoemission and provides detailed information on electron-electron and electron-phonon interactions in solids directly from time domain investigations with a resolution down to a few to tens of femtoseconds (1 fs = $10^{-15}$ s) [4,5,6,7].

The *e-ph* coupling strength and mean free path $l_{e-ph}$ in carbon nanotubes (CNTs) and in single-wall carbon nanotubes (SWNTs) is subject to continuing studies which try to determine the exact magnitude of $l_{e-ph}$. These efforts were stimulated partly by reported values of $l_{e-ph}$ exceeding 1 μm even at room temperature [8,9]. Here, an independent determination of the strength of the *e-ph* coupling would clearly be desirable. Time-resolved studies of carrier dynamics in graphite have already allowed a determination of the time-scale of electron-electron interactions as well as their energy dependence and provided first evidence of strong anisotropies of quasi-particle lifetimes [10,11]. Similar studies of *e-e* interactions in samples made of SWNT bundles [12,13] showed that interactions and relaxation times in SWNT samples are somewhat shorter, but qualitatively similar to those found in graphite. First time-domain studies on the strength of the *e-ph* coupling in SWNT bundles uncovered a weak *e-ph* coupling in metallic tubes which suggested correspondingly long *e-ph* scattering times [14,13]. In a promising study, another team has concentrated on relaxation dynamics in semiconducting tubes by transient absorption spectroscopy [15]. An unambiguous assignment of the observed transients to *e-ph* interactions, however, appears difficult.

Here, we have used femtosecond time-resolved photoemission to study the dependence of the *e-ph* energy transfer rate on electron and lattice temperatures. The lattice temperature was varied from 41 K to 320 K while the internal energy of the electronic system and the corresponding temperature reached values exceeding 1100 K. Using the theory by Allen [1] we derive the *e-ph* mass enhancement parameter $\lambda$ for SWNT bundles.

## 2. EXPERIMENT

SWNT samples (bucky-paper) were fabricated from commercial purified nanotube suspension (tubes@rice, Houston, Texas). These SWNTs, of typically 12 Å diameter, are arranged in a hexagonal close packed manner to form SWNT bundles with a few tens of nm diameter each. Since the samples have not been cut or treated with strongly oxidizing agents they reportedly contain only few opened SWNTs (o-SWNTs) and the majority should be closed (c-SWNTs) [16]. The bucky-paper sample was mounted back to back with an HOPG crystal onto a 1 cm diameter tantalum disk and could be heated or cooled from about 40 K up to 1200 K. In order to remove residual solvent and functional groups that may have been left from the purification procedure, SWNT samples were heated under ultra-high-vacuum conditions to peak temperatures of 1200 K.

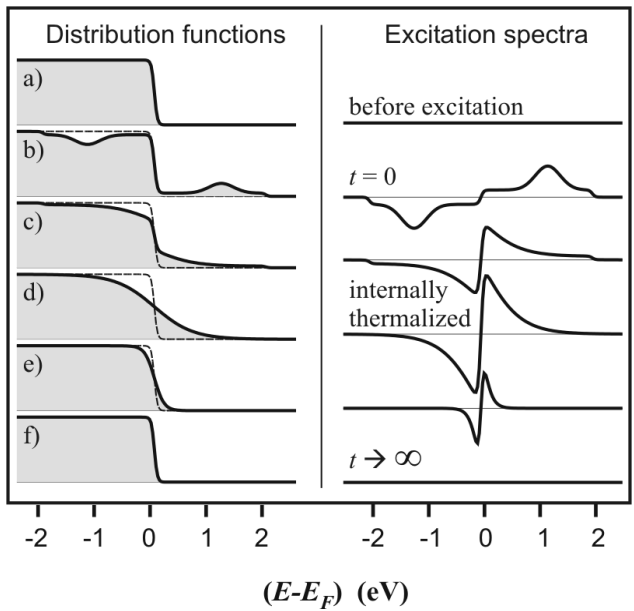

**Fig. 1**. Schematic illustration of the dynamics of a laser heated electron gas embedded in a comparatively cold lattice. The state of the electronic system is characterized by distribution functions (left) or alternatively excitation spectra (right) which directly show the changes in the distribution function at different stages of the thermalization process. a) The initially cold electronic system is b) excited optically and c) thermalizes internally before it cools back down to the lattice temperature d)-f).

Angle-integrated photoelectron spectra are recorded by the time-of-flight technique with a home-build spectrometer reaching an energy resolution of typically 10 meV at 1 eV kinetic energy. Ultrashort laser pulses are provided by a commercial femtosecond laser system consisting of TiSa oscillator, TiSa regenerative amplifier and an optical parametric amplifier (Coherent Inc.). The latter provides tuneable femtosecond laser pulses in the visible range. Pump pulses of 50-80 fs duration, at 2.32 eV photon energy with an estimated peak intensity of typically 50 µJ/cm$^2$ were focussed onto the sample for laser excitation and heating of electrons. A fraction of the pump beam was frequency doubled to generate 4.64 eV laser pulses (~5 µJ/cm$^2$) which were used to probe the electronic temperature by direct photoemission from the vicinity of the Fermi level. The probe pulse was directed onto the sample after a tuneable, well defined time-delay with respect to the pump pulse. The sample work function was found to be 4.52 eV ± 0.05 eV. More details on the experimental setup can be found elsewhere [13].

## 3. RESULTS AND DISCUSSION

### 3.1. Time-Resolved Photoelectron Spectroscopy

Photoelectron spectroscopy can be used to probe the temperature of a solid by a measurement of the width of the electron distribution near the Fermi-level. In combination with ultrashort laser pulses this technique allows to achieve a time-resolution down to a few tens of femtoseconds. The experimental concept of laser heating of an equilibrium electron distribution is illustrated schematically in Fig. 1. On the left side one finds the evolution of the distribution function $f^*(E)$ after laser excitation while the right side shows the so called excitation spectrum, *i.e.* the difference between the unperturbed and actual distribution function. Initially, a) the absorption of the visible laser pulse by the cold electron gas b) generates a non-equilibrium distribution which c) rapidly thermalizes and d) attains a new equilibrium at higher temperature within a few hundred femtoseconds. This process of internal thermalization is known to occur through rapid electron-electron (*e-e*) scattering and has been studied extensively [5,5]. Once the electron gas has thermalized at a higher temperature it subsequently d) & e) transfers its excess energy to the lattice by electron-phonon interactions which f) finally leads to a new equilibrium between lattice and electrons.

A set of experimental excitation spectra is shown in Fig. 2 as a function of the visible-pump UV-probe time-delay, which was varied from 0 to 5 ps. Evidently, the experimental spectra in Fig. 2 show the same characteristic features as discussed above. A Fermi-Dirac fit to the low energy end of the spectrum reveals the presence of non-thermalized electrons, in particular at higher energies and at small pump-probe time-delay. The characteristic time for thermalization after which the non-thermal contribution to photoelectron spectra becomes vanishingly small has previously been determined to be 0.2 ps [13]. As discussed above, one also finds that the temperature obtained from the FD fit continues to decrease from its maximum of 1100 K towards the lattice temperature ($T_l$ = 300 K). In the following we will describe how these spectra are used to quantify the rate of energy transfer from the electronic system to the lattice.

### 3.2. Data Analysis within the Two-Temperature Approximation

The coupling of a thermalized electron distribution at temperature $T_e$ with a host lattice at temperature $T_l$ can be described by a set of coupled differential equations:

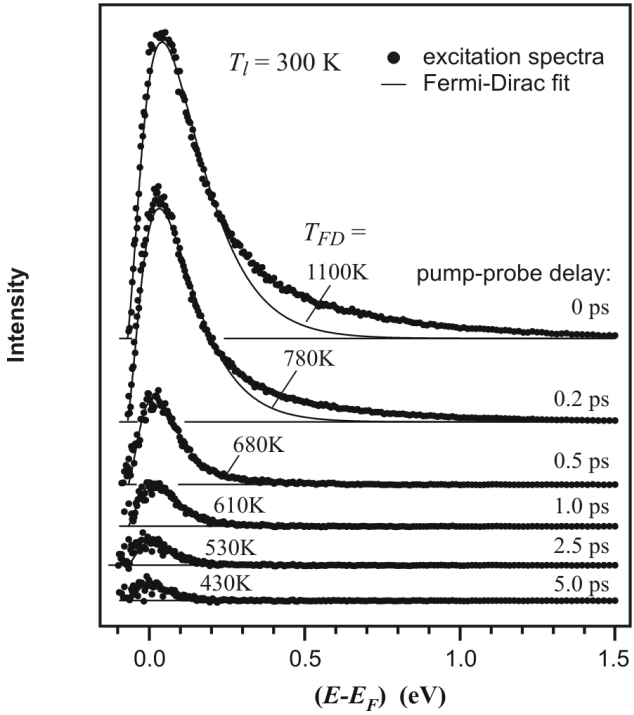

**Fig. 2**. Excitation spectra at varying pump-probe time delay from 0 to 5 ps. The thin solid line and temperatures are obtained from a fit of spectra to a Fermi-Dirac distribution.

$$C_e \frac{dT_e}{dt} = \nabla(k \nabla T_e) - H(T_e, T_l) + S(t) \quad (1)$$

$$C_l \frac{dT_l}{dt} = H(T_e, T_l) \quad (2)$$

This two-temperature model was first introduced by Anisimov *et al.* [17]. The source term $S(t)$ is given by the energy deposited into the electronic system due to absorption of the visible laser pulse and $k$ is the electronic heat diffusion coefficient. In SWNT samples – as for graphite – we can neglect the diffusion term perpendicular to the sample surface due to comparatively slow transport perpendicular to the tube axis and thus perpendicular to the bundles which generally lie in the surface. The terms on the left side quantify the rate of energy transfer from the electronic system to the lattice and thus include the corresponding heat capacities $C_e$ and $C_l$. Due to the fact that $C_e \ll C_l$ and due to slow electronic heat diffusion perpendicular to the bundle surfaces we can simplify these equations to obtain the electron-phonon coupling term $H(T_e,T_l)$ (in W/m³) after internal thermalization ($t > 0.2$ ps) from a single differential equation:

$$C_e \frac{dT_e}{dt} = -H(T_e, T_l) \quad (3)$$

This is also equivalent to the change of the internal energy of the electronic system.

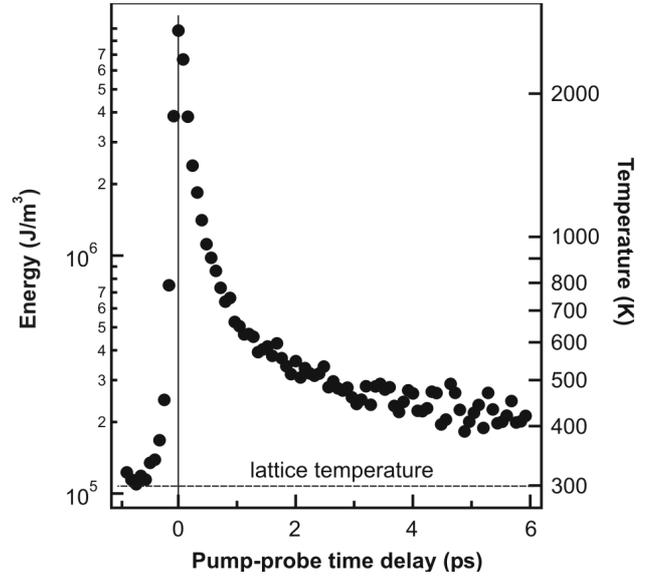

**Fig. 3**. Internal energy of the electronic system and the corresponding temperature as a function of the pump-probe delay.

$$H(T_e, T_l) = -\frac{dE_{int}}{dt} \quad (4)$$

The latter can be approximated using the energy weighted integral over experimental photoelectron intensities $I(E)$

$$E_{int} \propto 2 \int_0^\infty dE' \, E' \, I(E_{kin} + h\nu_{probe} - e\Phi) \quad (5)$$

where $h\nu_{probe}$ is the probe photon energy, $E_{kin}$ the electron kinetic energy with respect to the vacuum level and $e\Phi$ is the sample work-function. For simplicity, we assume that the spectrum directly reflects the electron distribution in the solid. This amounts to neglecting the energy dependence of photoemission matrix elements and seems justified considering the small energy region of interest. The proportionality constant of eq. (5) can be obtained by calibration of the calculated energy with the temperature and corresponding internal energy derived from the Fermi-Dirac fit at larger time delay, typically 3 ps.

The evolution of the internal energy of the electronic system as a function of time is plotted in Fig. 3. The high internal energy at early pump-probe delays would correspond to a temperature of approximately 2500 K which can be calculated using an electronic heat capacity $\gamma$ of 12 µJ/mole K² and a SWNT bundle density $r$ of 1.5 g/cm³ [13, 18]. Next, the data of Fig. 3 is used to obtain the energy transfer rate $H(T_e,T_l)$ by differentiation of the internal energy with respect to time (eq. (4)). The resulting electron to lattice energy transfer rate is plotted for three different lattice temperatures in Fig. 4. One finds that the data for different lattice temperatures almost coincide within the experimental scatter if plotted as a function of $(T_e - T_l)$, the temperature difference between electronic system and the lattice. The strongly nonlinear increase of $H$ in Fig. 4 shows that the electron to lattice energy transfer depends strongly on the electronic temperature $T_e$. The dependence on $T_l$ is not as clear because only three temperatures over a comparatively small temperature range, between 41 K and 320 K have been measured.

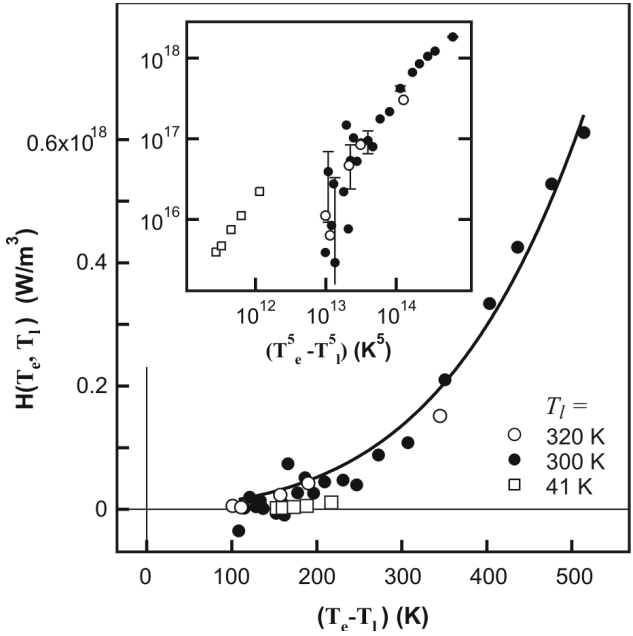

**Fig. 4**. Energy transfer rates calculated from curves as the one in Fig. 3 for three different lattice temperatures. The inset shows the same data as in the main plot but plotted versus $(T_e^5 - T_l^5)$ instead of $(T_e - T_l)$. The solid line is a fit to the 300 K data set using eq. 5.

### 3.3. Electron-Phonon Mass Enhancement and Scattering Times

The theory of Allen [1] now allows to relate the measured energy transfer rate to the electron-phonon mass enhancement parameter $\lambda$. In the low temperature limit, with Debye temperatures of the same order or higher than the electronic temperature, one expects a $T_e^5$ and $T_l^5$ power-law behaviour for the coupling term $H$ according to:

$$H(T_e, T_l) = h(T_e^5 - T_l^5) \quad (5)$$

with

$$h = \frac{144 z(5) k_B \gamma}{\pi \hbar} \frac{\lambda}{\Theta_D^2} \quad (6)$$

where $z(5) = 1.0369...$ is Riemanns Zeta function and $\gamma$ is the electronic heat capacity coefficient.

A best fit to the 300 K data in Fig. 4 (solid line and circles) yields $\lambda/\Theta_D^2 = (4 \pm 1) \cdot 10^{-10}$ K$^{-2}$ [19]. The low temperature, 41 K data set appears to be within the scatter of the 300 K and 320 K data sets but would lead to a somewhat larger value $\lambda/\Theta_D^2$ of $\sim 2 \cdot 10^{-9}$ K$^{-2}$ if analyzed independently. The latter data set, however, has somewhat less significance because only a few data points within a relatively small electronic temperature range are available. For a Debye temperature of about 1000 K [20] the fit to the 300 K data set gives $\lambda = (4 \pm 1) \cdot 10^{-4}$. Note, that higher values of the on-tube Debye temperature have been reported and that consequently, the e-ph mass enhancement term may be up to a factor of two larger. The value of $\lambda$ determined here is very small if compared, for example, with values found for noble metals $\lambda_{Cu} = 0.08$ and $\lambda_{Au} = 0.16$ or some metallic superconductors with $\lambda$ ranging up to 1.45 [3].

In the following we use the formulae by Allen [21] to estimate the e-ph scattering time:

$$\frac{1}{\tau_{e-ph}} = 24 \pi z(3) \frac{k_B}{\hbar} \frac{\lambda}{1+\lambda} \frac{T_e^3}{\Theta_D^2} \quad (7)$$

where $z(3) = 1.2020...$ . Fortunately, the scattering time is not affected by the choice of the Debye temperature as $\lambda/\Theta_D^2$ is determined directly from the fit to the experimental data. The 300 K data set gives an extraordinary long room temperature e-ph scattering time $\tau_{e-ph}$ of 15 ps while the low-T data would yield only 1.5 ps with – as explained above – somewhat less significance. Systematic errors due to day to day variations, for example, are difficult to quantify but may be within the range of the low- and room-temperature values. Using a group velocity of $10^6$ m/s the latter would correspond to a mean free path $l_{e-ph}$ of 15 µm. As pointed out previously [13], these numbers can be attributed exclusively to scattering in metallic tube species of the SWNT bundles since the electronic heat capacity – and thus the measured cooling process – is dominated by tubes with high density of states at the Fermi level.

### 4. CONCLUSIONS

The present time-domain study of electron-phonon interactions in metallic SWNTs in SWNT bundles revealed that the energy transfer rate between electrons and lattice is very small. The results suggest that room temperature scattering times $\tau_{e-ph}$ and mean free paths $l_{e-ph}$ can exceed 1.5 ps and 1.5 µm, respectively. The latter observation is in qualitative agreement with the results by Frank *et al.* [8] on MWNTs, Bachtold *et al.* [9] and recent transport studies by Appenzeller *et al.* [22,23] on SWNTs, which all estimate e-ph mean free path to significantly exceed 1 µm at room temperature. The experiments furthermore demonstrate that energy transfer between the electronic system and its host lattice depends strongly, *i.e.* approximately with the 5$^{th}$ power on electronic and lattice temperatures.

**Acknoweldgement**: R.F. acknowledges financial support by the Alexander von Humboldt foundation through a Humboldt Fellowship. It is our pleasure to acknowledge generous support by G. Ertl.